# Crossover from coherent to incoherent scattering in spin-orbit dominated $Sr_2IrO_4$

Mehmet Fatih Cetin[1], Peter Lemmens[1], Vladimir Gnezdilov[1,2], Dirk Wulferding[1], Dirk Menzel[1], Tomohiro Takayama[3], Kei Ohashi[3], and Hidenori Takagi[3,4]

[1]*Institute for Condensed Matter Physics, Technical University of Braunschweig, D-38106, Germany und Niedersächsiche Technische Hochschule, Germany*

[2]*B.I. Verkin Institute for Low Temperature Physics and Engineering, NASU, UA-61103 Kharkov, Ukraine*

[3]*Dept. Advanced Material Science, University of Tokyo, Kashiwa 277-8651, Japan*

[4]*RIKEN Advanced Science Institute, Saitama 351-0198, Japan*

**Abstract**

Strong spin-orbit interaction in the two dimensional compound $Sr_2IrO_4$ leads to the formation of $J_{eff}=1/2$ isospins with unprecedented dynamics. In Raman scattering a continuum attributed to double spin scattering is observed. With higher excitation energy of the incident Laser this signal crosses over to an incoherent background. The characteristic energy scale of this cross over is identical to that of intensity resonance effects in phonon scattering. It is related to exciton-like orbital excitations that are also evident in resonant X-Ray scattering. The crossover and evolution of incoherent excitations are proposed to be due to their coupling to spin excitations. This signals a spin-orbit induced entanglement of spin, lattice and charge degrees of freedoms in $Sr_2IrO_4$.

# I. Introduction

In correlated electron systems novel quantum phases and properties result very often from an uncommon hierarchy of energy scales. Recently interesting cases concern spin-orbit coupling (SOC) comparable or even superior to Coulomb energies [1]. Spin-orbit coupling is relevant for some 4d and 5d systems exemplified by the iridates $Na_4Ir_3O_8$ and $Na_2IrO_3$ with a spin liquid [2] and a topological insulating [3] state, respectively, and $Sr_2IrO_4$ as a special Mott insulator [4]. In strongly correlated 3d electron systems SOC is usually comparably small leading to moderate changes of the magnetic excitations by anisotropy gaps in the long range order state. With only few exceptions magnetic modes are strictly decoupled from charge excitations. With strong SOC, however, the spin excitation spectra differ considerably from the case of 3d systems. This is due to topological effects as well as the mixing with orbital and charge excitations. It is presently a matter of debate which consequences this overlap and interaction has for charge excitations close to the optical gap (Mott-Hubbard) gap, e.g. the existence of unconventional superconductivity and topological modes is discussed.

The structure of $Sr_2IrO_4$ is tetragonal and known from some cuprates, with magnetic ions ($Ir^{4+}$, $5d^5$) on a square plane connected by oxygen ions forming octahedra. The compound shows a concomitant antiferromagntic (AFM) order with canted, in-plane, ferromagnetic moments (0.14$\mu_B$ per Ir) and structural distortions with in-plane rotations of the $IrO_6$ octahedra enlarging the unit cell for $T<T_N=240K$ [5]. The rotations and the moments are directly related via SOC [6]. For lower temperatures further anomalies are observed, e.g. an enhanced electric permittivity at around 100 K which is attributed to modulations of the Ir-O-Ir bond angle [7].

The $Ir^{4+}$ low lying $t_{2g}$ states form lower $J_{eff}=3/2$ and higher energy $J_{eff}=1/2$ isospin states with a splitting given by the large SOC coupling constant $\lambda \approx 0.4eV$ [8]. With 5 electrons this situation corresponds to a single band with $J_{eff}=1/2$ [4]. More conventional 5d compounds have a considerable metallicity. This signals the unusual Mott state with relevance of electronic correlations for the narrow, spin-orbit induced $J_{eff}=1/2$ states [6].

Using the phase sensitivity of resonant inelastic X-Ray scattering (RIXS) the $J_{eff}$ level scheme has been directly probed [9]. Interband transitions across the Mott gap and charge-transfer excitations from the low energy O $2p$ band to the Ir $5d$ $t_{2g}$ bands (upper Hubbard band) and to the Ir $e_g$ $3z^2-r^2$ states show up with characteristic energies of 0.5, 2.5, 3.2, and 6.0 eV [10]. These modes are coherent excitations with a very small dispersion of the Mott gap in contrast to theory [1]. More recently a strongly dispersing, lower energy mode (0.4-0.8eV) has been observed and related to spin-orbit excitons of the $J_{eff}$ states [11].

The above mentioned energies are in good agreement with optical absorption studies [12]. The optical absorption renormalizes gradually with increasing temperatures which is interpreted as a reduction and filling up of Mott-Hubbard with temperature. Optics also shows phonon anomalies, e.g. a 3% softening of a phonon mode at 660 cm$^{-1}$ that modulates the Ir-O-Ir bond angle [12]. A list of these phonons is given in Table 1 together with Raman active modes.

For a deeper understanding of this system it is of pivotal importance to probe both lattice and electronic degrees of freedom using suitable spectroscopic techniques. Questions to be addressed concern the typical energies and character of the excitations and whether temperature can be used to tailor the system from an anisotropic to an isotropic magnetism with possible novel excitations.

Inelastic light scattering (Raman scattering) probes lattice degrees of freedom as well as electronic correlations via a virtual electron-hole pair. The characteristic energy of this state can be tuned into resonance with the systems characteristic energies. To our knowledge such resonant optical experiments have not been performed on isospin states up to now. Our data indeed show a pronounced resonance of phonon as well as high energy scattering with the energy of the incident Laser. We observe a crossover from coherent (Raman) to incoherent scattering (fluorescent-like) with a similar resonance energy. We attribute this crossover to the excitation of an orbital-exciton that decays into spin excitations. The observed dynamics, e.g. also the intensity as function of temperature, is entirely different from correlated 3d electron systems.

## II. Experimental details

$Sr_2IrO_4$ single crystals with approximate dimension of 1 x 0.5 x 0.5 mm$^3$ were grown by using a flux method. Crystals were cleaved to establish a virgin, optically flat surface and oriented with Laue diffractometry. A later cleaning of these surfaces using acetone or methanol is not advisable due to its detoriating effect on the spectra. Raman spectroscopic studies were performed in quasi-backscattering geometry in *ab* and *c*-planes with a solid-state laser (λ=532 nm and 4 mW laser power) as excitation light source. The polarizations (xx) and (yx) denotes parallel and crossed polarization of incident and scattered light, respectively. (y'x') indicates a rotation of the crystal by 45 ° along c axis and ''u'' stands for unpolarized.

The crystal structure of $Sr_2IrO_4$ is $D_{4h}^{20}$ – $I4_1/acd$, Nr. 142 [5] and is derived from a distorted $K_2NiF_4$ (I4/mmm)-type tetragonal structure. It belongs to the tetragonal - ditetragonal dipyramidal class (4/m 2/m 2/m), with four 2-fold and one 4-fold axes including a center of inversion. Thirty-two Raman active modes are expected according to a factor group analysis [13], $\Gamma_{Raman}=4A_{1g}+ 7B_{1g} + 5B_{2g} + 16E_g$. The different ions of $Sr_2IrO_4$ contribute to all symmetries with the exception of Ir that does not contribute to $A_{1g}$ symmetry. The $K_2NiF_4$ (I4/mmm) structure yields only four modes, $\Gamma_{Raman}= 2A_{1g} + 2E_g$.

Measurements were carried out in an evacuated, closed-cycle cryostat. The spectra were collected via a triple spectrometer (Dilor-XY-500) and a micro Raman setup (Horiba Labram) equipped with liquid nitrogen cooled charge coupled device (CCD) detector. The measurements of the low (high) energy modes have been performed with different spectral resolution of 3-4 cm$^{-1}$ and 10 cm$^{-1}$, respectively, using a slit width of d=100 (250) μm. For resonance Raman investigations different ion gas and solid state Lasers with excitation lines (2.2-2.7 eV, 632-457nm) were applied using a power level of P=4 mW. Measurements using a He-Cd Laser (325nm) have been performed with the micro Raman setup.

## III. Results

In Figure 1 Raman spectra of $Sr_2IrO_4$ in different polarizations are shown, including in-plane as well as out-of-plane polarizations. In Figure 1(a) six pronounced and five weaker modes are observed, in good agreement with the allowed $4A_{1g}+ 7B_{1g}$ modes. For details we refer to Table 1. We detect four out of five expected $B_{2g}$ modes in (yx) polarization. In (y'x') polarization four out of seven $B_{1g}$ modes are observed, see Figure 1(b).

With incident light polarized parallel to the c axis and no analyzer of the scattered light, denoted by (zu), we detect four of four allowed $A_{1g}$ modes, see Figure 1(c). We detect none of the sixteen $E_g$ modes in crossed polarization. This could be attributed to small scattering cross section. It is obvious from these data that in each polarization a different mode has a dominating intensity, i.e. at 277 cm$^{-1}$, 390 cm$^{-1}$ and 560 cm$^{-1}$ modes in (xx), (yx) and (zu) polarizations, respectively. This is evidence for a pronounced spatial anisotropic polarizability as expected for a layered material. A mode assignment based on the mode frequency and a comparison with earlier data from IR experiments [12] is given in Table 1.

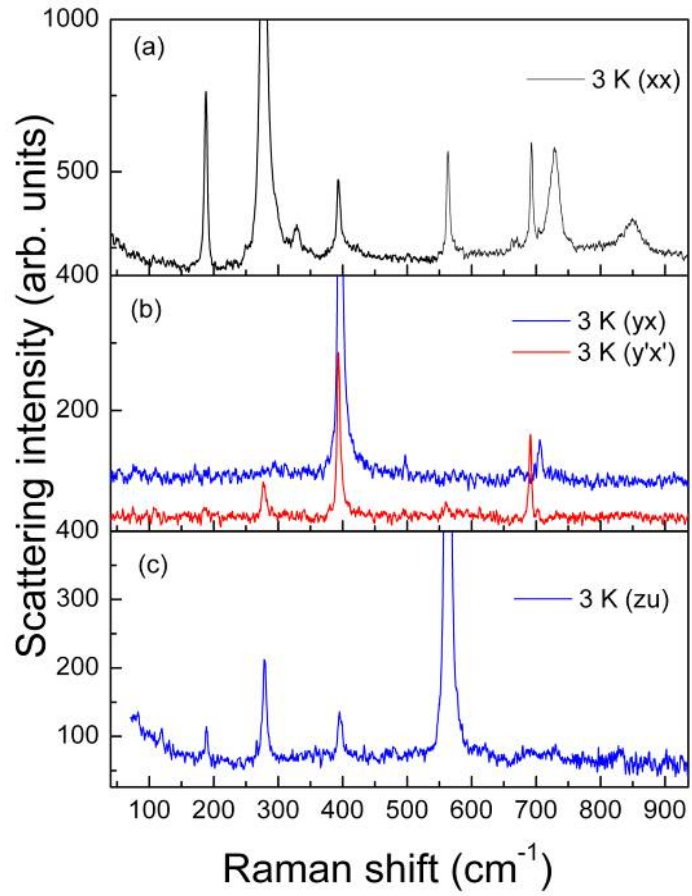

FIG. 1. Polarization dependence of the Raman spectra measured at 3 K in (xx) polarization (a), (yx) and (y'x') polarizations (b), and (zu) polarization (c).

Table 1. Raman and infrared [12] active modes observed in $Sr_2IrO_4$.

| Frequency (cm$^{-1}$) | Polarization | | | | Assignment | Comments |
|---|---|---|---|---|---|---|
| | xx | yx | y'x' | zu | | |
| 98 | IR | | | | $E_u$ (external modes, Sr to $IrO_6$) | - |
| 112 | IR | | | | | |
| 138 | IR | | | | | |
| 187 | + | - | - | + | $A_{1g}$ (Sr against $IrO_6$) | - |
| 252 | + | - | - | - | $B_{1g}$ (Sr) | - |
| 270 | IR | | | | $A_{2u}$ (Ir-O-Ir bond angle) | softening (2 %) |
| 277 | + | - | + | + | $A_{1g}$ (Ir-O-Ir bending) | anomalous softening (6 %) |
| 327 | + | - | - | - | $A_{1g}$ (Oxygen) | vanish at HT |
| 365 | IR | | | | $A_{2u}$ (Ir-O-Ir bond angle) | anomalous softening (3 %) |
| 390 | + | - | - | - | $B_{1g}$ (Ir) | - |
| 497 | - | + | - | - | $B_{2g}$ | dominating intensity |
| 560 | + | - | + | + | $B_{1g}$ (Oxygen) | asymmetric line shape |
| 660 | IR | | | | $A_{2u}$ (Ir-O bond) | broadening, softening |
| 666 | + | - | - | - | $B_{1g}$ (Oxygen) | very weak |
| 690 | + | - | + | - | $B_{1g}$ (Oxygen) | - |
| 706 | - | + | - | - | $B_{2g}$ (Oxygen) | - |
| 728 | + | - | - | - | $B_{1g}$ (Oxygen, breathing) | anomalous broadening |
| 850 | + | - | - | - | $B_{1g}$ (Oxygen) | very broad, asymmetric line shape |
| 1240 | + | - | - | - | 2-Phonon of 666 cm$^{-1}$ | |
| 1345 | + | - | - | - | 2-Phonon of 690 cm$^{-1}$ | |
| 1467 | + | - | - | - | 2-Phonon of 728 cm$^{-1}$ | dominating with low energy Lasers |
| 1800 | + | - | - | - | 2-Magnon / electronic | large linewidth, vanishing at HT |

A comparison of spectra at two temperatures within the low, medium and high spectral ranges is depicted in Figures 2(a), 2(b) and 2(c), respectively. All observed phonons increase in intensity and get sharper in line shape at low temperatures. The lower energy $A_{1g}$ modes, 187 cm$^{-1}$ and 277 cm$^{-1}$, can be assigned to Sr stretching against IrO$_6$ and the important bending mode of the Ir-O-Ir bonds, respectively [14,15]. Weak modes at 252 cm$^{-1}$ and 327 cm$^{-1}$ vanish at high temperature. Breathing modes of apical oxygen showing up in the frequency range of 500 cm$^{-1}$ to 850 cm$^{-1}$ are asymmetric and have Fano-like line shapes. This is probably related to an interaction with some continuum of excitations.

In the higher frequency range we observe additional maxima in the scattering intensity. These modes at around 1240 cm$^{-1}$, 1345 cm$^{-1}$ and 1467 cm$^{-1}$ have a very large linewidth (~50 cm$^{-1}$) and look more like density of states than usual phonon modes. Therefore we assign them to 2-phonon scattering of the 666 cm$^{-1}$, 690 cm$^{-1}$ and 728 cm$^{-1}$ phonon modes, respectively. The observation of 2-phonon scattering with large intensity resembles to observations in correlated 3d electron systems with orbital dynamics [16,17].

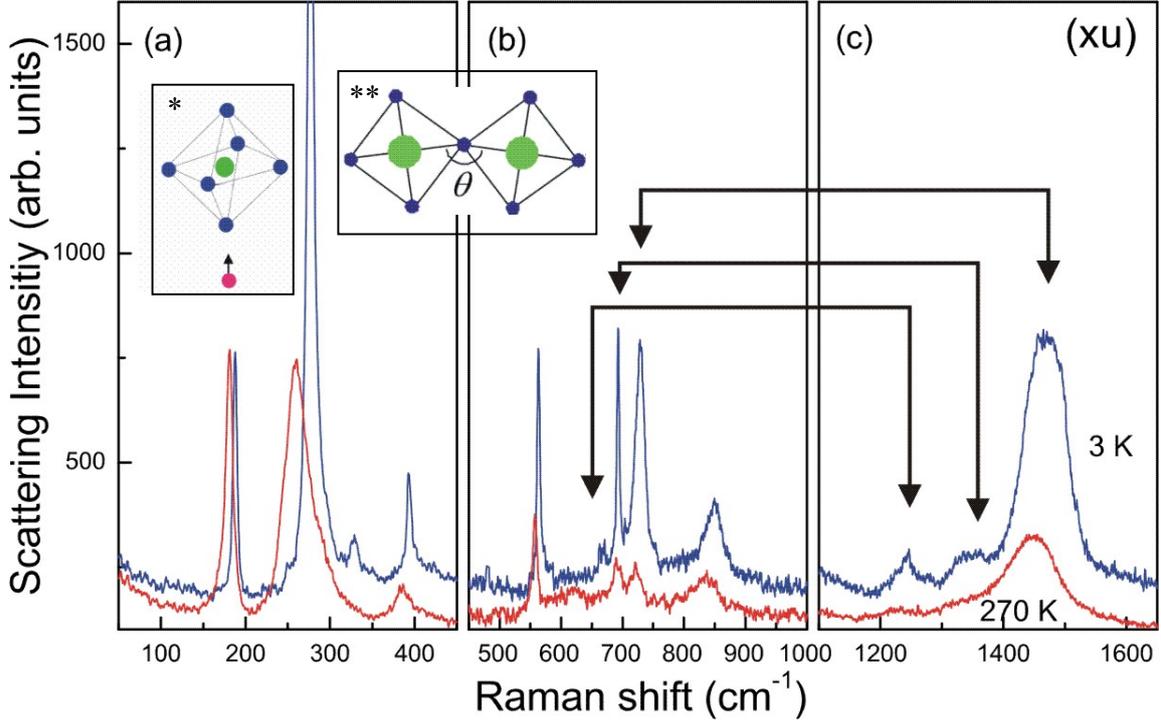

FIG. 2. Comparison of Raman spectra at 3 K (blue) and at 270 K (red) in the (a) low, (b) medium, and (c) high energy spectral ranges in (xu) light scattering polarization. *Displacement pattern of $A_{1g}$ mode at 187 cm$^{-1}$ (Sr vs. IrO$_6$), **Modulation of Ir-O-Ir bond angle resulting in $A_{1g}$ mode at 277 cm$^{-1}$.

The normalized intensity, linewidth (FWHM) and frequency of particular modes which behave anomalously are depicted in Figure 3. All modes display a general increase in intensity and a decrease in linewidth as temperature decreases. The phonon at 277 cm$^{-1}$ shows an anomalously large shift (20 cm$^{-1}$ ~ 6 %). It is assigned to a modulation of the Ir-O-Ir bond angle. The linewidth of the mode at 728 cm$^{-1}$ shows a kink at approximately 100 K. For temperatures above 100 K, its linewidth broadens substantially with a saturation for T>$T_N$=240 K. It is assigned to a breathing mode of the apical oxygen. Frequency and intensity of the phonon and 2-phonon mode (728 cm$^{-1}$ and 1467 cm$^{-1}$) modes behave similarly with increasing temperatures.

In Sr$_2$IrO$_4$ as well as other oxide based correlated electron systems the weakly hybridized states of the transition metal oxide coordinations form narrow bands with a highly structured density of states. If these energy scales fit to the photon energies of incident Laser radiation resonance Raman Scattering (RRS) can be observed. It is noteworthy that the Raman excitations process couples the induced electron hole pair exactly to the states that are relevant for magnetism and spin phonon coupling and can give intriguing information. Such experiments on two magnon states have supported the Hubbard model description of high temperature superconductors [18].

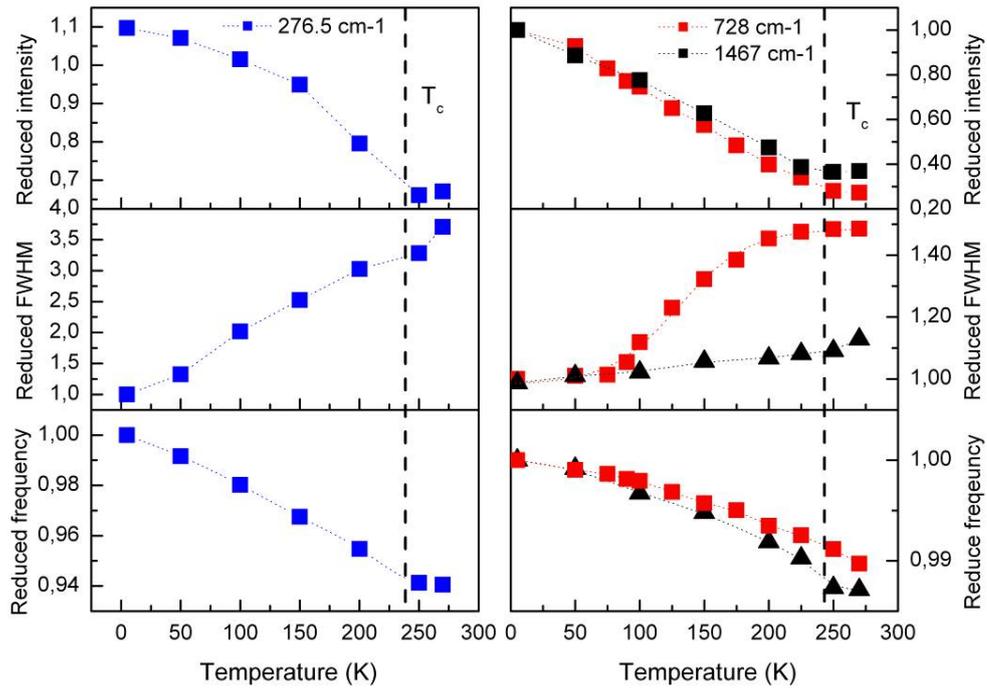

FIG. 3. Temperature dependence of the normalized integrated intensities, linewidth (FWHM) and peak positions of the phonon modes at 187, 277, 728, and the 2-phonon mode at 1467 cm$^{-1}$.

In Figure 4 (a) we show the dependence of selected phonon intensities (187 cm$^{-1}$, 277 cm$^{-1}$) and the 2-phonon (1467 cm$^{-1}$) intensity on the incident laser energy. There exist a very sharp maximum at 2.5 eV. This characteristic energy is also known from RIXS and IR absorption. For the mode at 728 cm$^{-1}$ and 395 cm$^{-1}$ there is a broader profile and a resonance to higher energies, respectively, see Figure 4 (b). This is attributed to different electronic states relevant for these phonon modes. Furthermore, also the intensity ratio of the 2- phonon to the 1-phonon modes shows a maximum at 2.58 eV, see Figure 4 (c).

This is evidence that the enhanced intensity of 2-phonon scattering is also related to a resonant scattering process. For HTSC and manganites such pronounced resonances of phonon scattering intensities have not been observed.

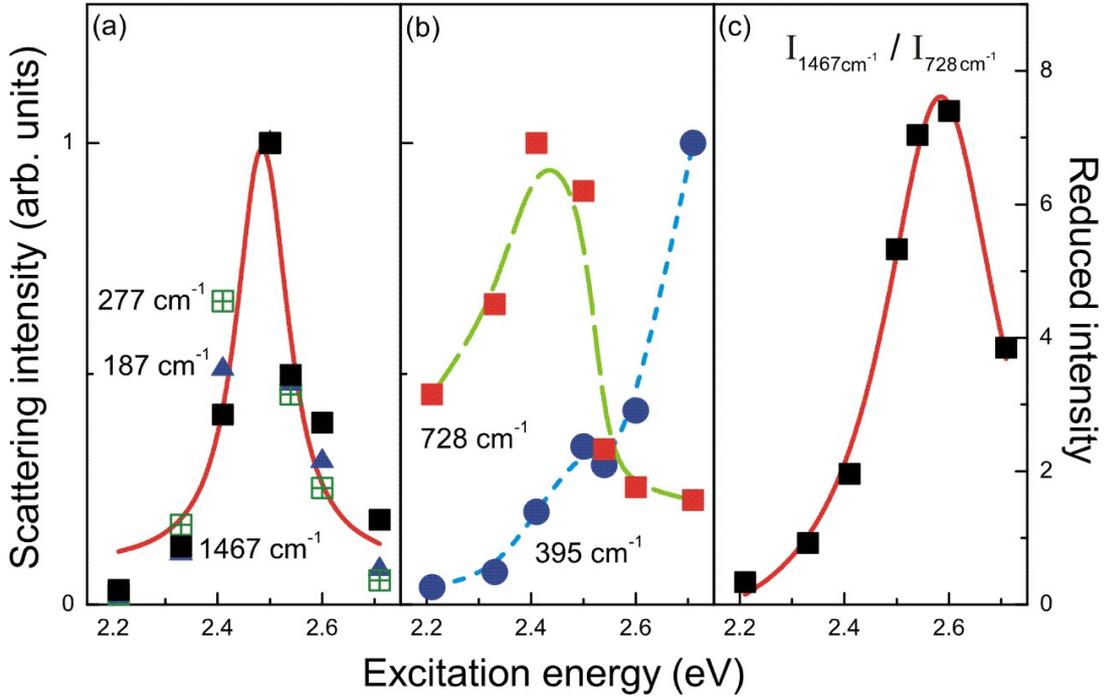

FIG. 4. a) Intensity of phonon Raman scattering as function of Laser excitation energy. a) Phonon intensity at 187 cm$^{-1}$, 277 cm$^{-1}$, and the 2-phonon intensity at 1467 cm$^{-1}$; (b) Phonon intensity at 395 cm$^{-1}$ and 728 cm$^{-1}$; (c) intensity ratio of 2-phonon (1467 cm$^{-1}$) to 1-phonon scattering (728 cm$^{-1}$). The solid red line is a fit to a Lorentzian line shape.

We observe a further very broad maximum (width 1000 cm$^{-1}$) at low temperatures and centered at higher energy (1800 cm$^{-1}$), see Figure 5. This maximum is a candidate for two spin excitations as the corresponding branches in RIXS have been observed with typical energies at 800 and 1600 cm$^{-1}$ in dependence of wave vector and a similar large dispersion [11]. It should be noted, however, that despite similarities of the $J_{eff}=1/2$ to a $s=1/2$ quantum spin system the dynamics of the former state can be very different. This concerns the energy, lineshape and temperature dependence of two particle processes and has its origin in the different dependence of isospin exchange processes on additional orbital and geometric factors. To our knowledge no experimental or theoretical investigation of isospin excitations in Raman scattering exists.

From Figure 5 we also deduce a strong suppression of the integrated scattering intensity with increasing temperatures and its vanishing at approximately 250 K. Such a temperature induced suppression is not observed for s=1/2 two-magnon scattering even for T≈$T_N$. Two magnon scattering is based on a local spin exchange and their characteristic energies depend only weakly on long range ordering.

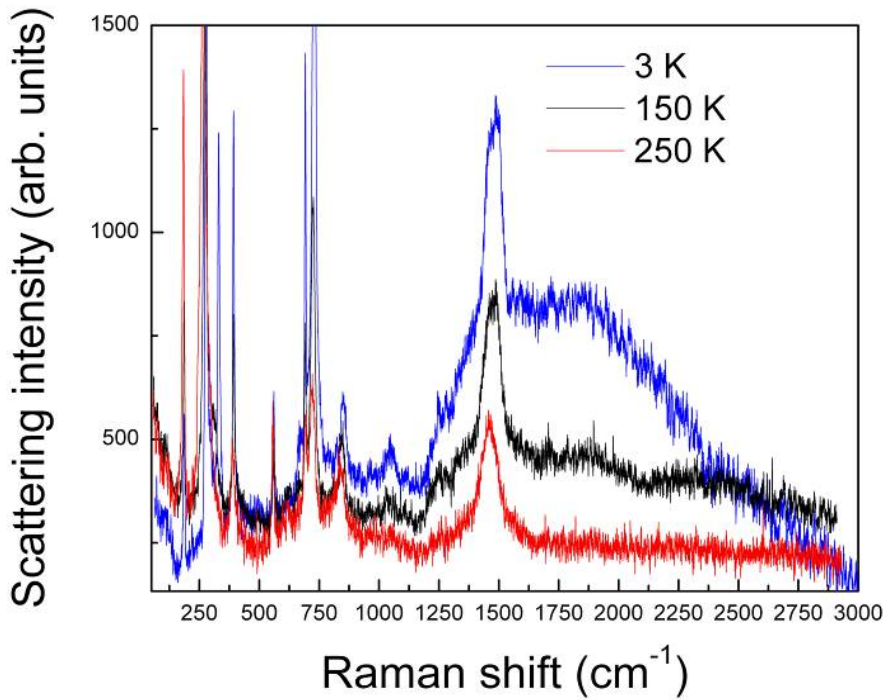

FIG. 5. High frequency spectral range with a broad asymmetric maximum. The measurements are performed in (xu) polarization with a 561-nm excitation at T=3 K, 150 K, and 250 K.

Increasing the Laser energy above 2.6 eV ~ 476 nm the broad Raman mode is replaced by an even broader background that seemingly shifts to higher energies, see Figure 6. In contrast to the former Raman process attributed to a two particle excitation of coherent (magnon-like) excitations the mode observed with higher energy is incoherent and shifts only on the relative scale "Raman shift". In the inset of Figure 7 this signal is plotted on an absolute energy scale. We find a negligible frequency shift as function of the incident Laser energy. The onset of the second process is around 2.55 eV, an energy which is again related to the characteristic resonance in $Sr_2IrO_4$. As in previous RIXS experiments [11] this energy has been related to the creation of excitons in the $J_{eff}$=1/2 to 3/2 isospin level manifold, we denote this regime *exciton regime*.

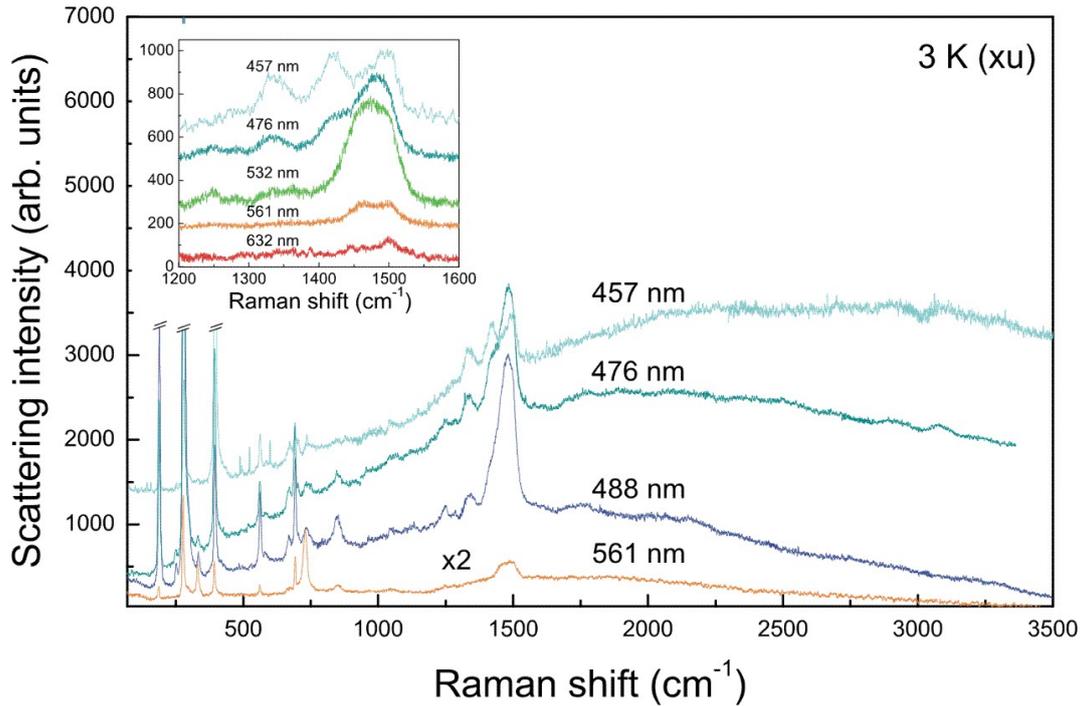

FIG. 6. Resonance Raman spectra with increasing excitation energy at T=3 K in (xu) polarization. The scattering shows a crossover from coherent Raman scattering with an incident laser wavelength of 561 nm to an incoherent, fluorescent like signal with higher excitation energy. The inset shows the frequency regime of two phonon scattering with increasing excitation energy from 632 nm to 457 nm, from bottom to top. These curves are shifted for clarity.

In the exciton regime there is also a drastic change of the 2 phonon lineshape evident around 1500 cm$^{-1}$, see inset of Fig. 6. With lower excitation energy only a single, hump-like mode is observed. Such a line shape origins from multiparticle scattering with no preferred scattering vector. With higher energy this signal is strongly enhanced and splits into a sequence of peaks. Its lineshape is rather similar to observations in manganites [16] or TiOCl [17] with orbital ordering and incommensurate distortions. In TiOCl the 2 phonon scattering evolves at the lower boundary of the energy scale of two-particle electronic/magnetic scattering. A similar coincidence of energy scales is exists in $Sr_2IrO_4$. The main difference between the 3d and the 5d compound, however, is the control of scattering lineshape and intensity by the excitation energy. In the former compounds such resonant processes with a cross over do not exit.

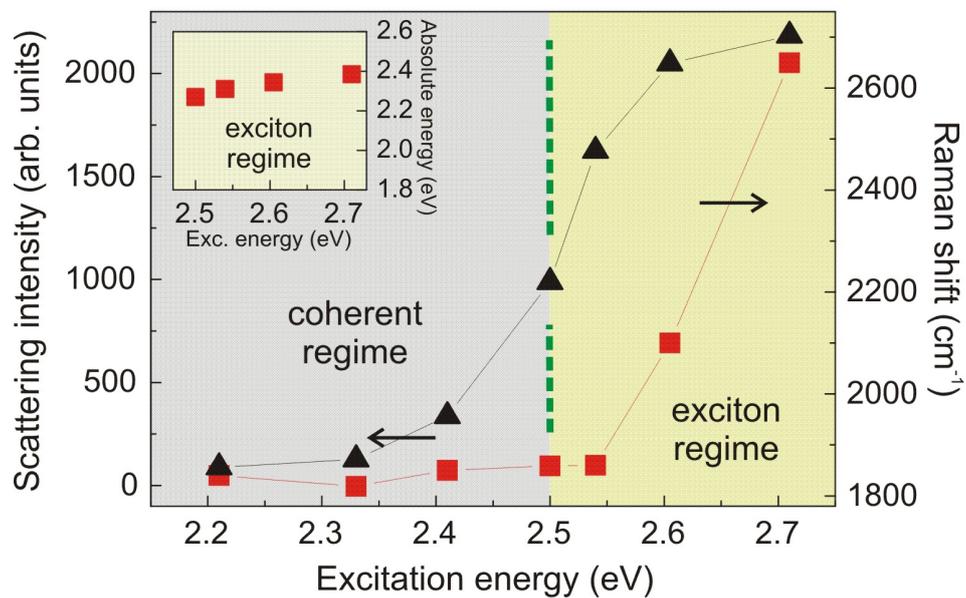

FIG. 7. Intensity and Raman shift of the broad high energy mode as function of Laser excitation energy. The dashed line marks the crossover from a coherent regime to an exciton regime. Inset: In the exciton regime the absolute energy of this mode does only weakly depend on excitation energy.

# IV. Discussions

Iridium oxides are presently intensively studied to understand the interplay of strong SOC with correlation effects, not known from 3d or 4d transition metal oxides. This interplay is based by one part to a different hierarchy of energy scales and by another part on symmetry aspects of SOC. The latter may lead to, e.g. topological phases/excitations and enhanced spin-phonon coupling.

The unusual physics of $Sr_2IrO_4$ has two previously known fingerprints, small canted moments for $T<T_N$ that are related to the Mott insulating state as well as pronounced electronic resonances. Only recently it has been found that these resonances consist of a group of dispersing/coherent spin as well as coherent and incoherent electronic modes that nearly overlap. The electronic modes consist of spin-orbit excitons and an electron hole continuum, respectively, and are based on an isospin $J_{eff}=1/2$ manifold of states [11].

Raman scattering has the potential to investigate all aspects of this complex dynamics. Despite its very high sensitivity, it lacks momentum resolution. Therefore we highlight the effect of incident Laser energy on the scattering linewith and intensity. Resonance Raman scattering is observed in many transition metal oxides as an intensity variation. However, a change of linewidth is a rare phenomenon. In the following we will first discuss the phonon anomalies and their resonance effects and then the properties of the high energy magnetic / electronic scattering.

All phonons show an increase of intensity with decreasing temperature. We assign this seemingly trivial effect to the changing optical penetration depth and scattering volume. The optical (Mott-Hubard) gap in $Sr_2IrO_4$ is small and the states in the gap contributing to the optical conductivity and penetration depth are only gradually depleted with decreasing temperature. More specific and larger anomalies concern modes that are related to Ir-O bending and stretching modes. In general these anomalies are noteworthy, as due to their low energy the O2p states only weakly hybridize with Ir 4d states. Therefore one expects small related spin phonon coupling constants. In contrast, the observed anomalies are rather strong, see, e.g. the 6-7% softening of the 277-cm$^{-1}$ mode. Therefore we have to assign them to different coupling mechanisms. The gradual evolution of the phonon frequencies and only moderate changes at the ordering temperature point to a phonon coupling to the SOC induced gap itself. As the magnitude of the gap is due to an interplay of SOC with correlations the dependence of SOC on local symmetry leads to an effective electron-phonon coupling mechanism. This is in agreement with Table 1 where phonons related to the Ir – O bonding angle and length show appreciable anomalies without being very specific. Therefore, the linewidth of these modes and the intensity of multiphonon scattering is not dominated by conventional anharmonicity via phonon scattering but by excitations of the isospin manifold of states. The abrupt change of the multiphonon lineshape with excitation energy shown in the inset of Figure 6 is thereby taken as an indication of the entanglement of structural and electronic degrees of freedom in this compound.

The phonon intensities in Fig. 4 show two resonance energies at 2.5 eV and above 3eV. These anomalies point to the same electron phonon coupling as the characteristic energies that are also resolved in optics and RIXS are due to the isospin state. In optics the 2.5-eV contribution shows a weak temperature dependence [12]. With decreasing temperatures there is a shift of spectral weight from around 0.5eV to this higher spectral range. The lower energy regime corresponds most probably to excitations across the Hubbard bands that form the optical gap. This spectral weight shift is very probably the origin for most of the phonon anomalies as function of temperature shown in Fig. 3.

While the intensity of phonon scattering is a rather sharp marker of the correlated electron energy scales their frequency dependence is not characteristic as it is mainly determined by other electronic states. In contrast magnetic and electronic Raman scattering are directly related to the relevant energy scales. Prominent examples are the parent compounds of HTSC that show two magnon scattering at approximately *3J*, with *J* the exchange coupling, [18], and the spin liquid state on the Kagome lattice with a broad response distributed from zero energy to at least *5-6 J* [19]. In Sr$_2$IrO$_4$ the broad asymmetric feature centered at around 1800 cm$^{-1}$ (~0.23 eV) corresponds to an exchange coupling J=850 K. Here we use a 2D bond model with 4 nearest neighbors and a spin of ½, i.e. $E_{max}=J(2zs-1)$. This energy lies a bit below the range of earlier estimations of J=1000 K [20].

There are several properties that are uncommon compared to two magnon scattering. All of them speak individually for an interaction of the spin system with other degrees of freedom. The normalized linewith of approximately 1000 cm$^{-1}$ (w=$\Delta E/E_{max}$=0.55) is very large compared to NiF$_2$, a canted AF, (w=0.076) [21] and still larger than in the parent compounds of high temperature superconductors (w=0.37) [18]. It is, however, quite similar to the width (0.11 eV ~ 900 cm$^{-1}$) of the resonance profile centered at 2.5eV, which would evidence a decay via charge excitations.

The magnetic scattering intensity in Sr$_2$IrO$_4$ is rapidly depressed with increasing temperatures. This temperature dependence is paralleled by the above mentioned spectral weight shift from 0.5 to 2.5eV in optics and is not expected in a conventional system without orbital degrees of freedom.

Finally we mention the rather special suppression of the two-magnon scattering with increasing excitation energy and crossover to an incoherent broad signal. This crossover is attributed to the proximity of the spin excitations to charge modes at the top of the Mott-Hubbard gap. This intermixing in the exciton regime is a direct consequence of the dominating spin-orbit coupling. In contrast to the sharp resonance of the phonon modes the maximum in the incoherent exciton regime is extremely broad. This could be due to metastable electronic configurations like a spin/charge polarons that are induced by the optical absorption process with energies above the resonance, i.e. threshold energy. The formation of polarons has been shown to induced resonant Raman scattering, e.g. in the double perovskite Ba$_2$MnWO$_6$ [22]. However there are also other candidates as long living topological excitations or solitons [23].

## V. Conclusions

In summary we have presented resonance Raman data of a SOC dominated Mott insulator. Due to entanglement of geometry, spin and orbital degree of freedoms, this $J_{eff}=1/2$ system shows novel spectroscopic effects. There is a rather sharp resonance of the phonon intensity at 2.5eV and other anomalies due to an O2p – Ir $t_{2g}$ excitation across the Mott-Hubbard gap. The observed temperature dependence is attributed to a spectral weight shift induced by the lowest electronic levels in the optical gap at 0.5 eV. With an Laser excitation energy above the resonant threshold energy the magnon scattering crosses over to a broad incoherent scattering. We attribute this to a coupling to charge modes and the existence of an exciton regime. The presented data demonstrate an interesting case of a correlated electron system dominated by spin-orbit coupling.

## Acknowledgements

We thank G. Jackeli, and K.-Y. Choi for important discussions. This work was supported by NTH, B-IGSM of the TU-BS and DFG.